\documentclass[preprint,authoryear,12pt]{elsarticle}

 \usepackage{graphicx}
\usepackage{epstopdf}
\usepackage{amsmath}
\usepackage{epsfig}
\usepackage{amssymb}
\DeclareMathOperator{\sech}{sech}

\usepackage[ps2pdf,%
a4paper=true,%
breaklinks=true,%
colorlinks=true,%
pdfauthor={First Author et al.},%
pdftitle={Template for manuscripts in Advances in Space Research}%
]{hyperref}

\journal{Advances in Space Research}

\begin{document}

\begin{frontmatter}



\title{Dispersive shock waves in partially ionised plasmas}


\author{I. Ballai}\corref{cor}
\address{Plasma Dynamics Group, University of Sheffield, Hounsfield Road, Hicks Building, Sheffield, S3 7RH, United Kingdom}
\cortext[cor]{Corresponding author}
\ead{i.ballai@sheffield.ac.uk}

\author{E. Forg\'acs-Dajka}
\address{Department of Astronomy, E\"otv\"os University, P\'azm\'any P\'eter s\'et\'any. 1/A, H-1118 Budapest, Hungary}
\ead{e.forgacs-dajka@astro.elte.hu}

\author{A. Marcu}
\address{Babes-Bolyai University, Cluj-Napoca, str. Mihail Kogalniceanu nr. 1, 3400, Cluj-Napoca, Romania}
\ead{ucramidnas@gmail.com}

\begin{abstract}

Compressional waves propagating in the partially ionised solar lower atmospheric plasmas can easily steepen into nonlinear waves, including shocks. Here we investigate the effect of weak dispersion generated by Hall currents perpendicular to the ambient magnetic field on the characteristics of shock waves. Our study will also focus on the interplay between weak dispersion and partial ionisation of the plasma. Using a multiple scale technique we derive the governing equation in the form of a Korteweg-de Vries-Burgers equation. The effect of weak dispersion on shock waves is obtained using a perturbation technique. The secular behaviour of second order terms is addressed with the help of a renormalisation technique. Our results show that dispersion modifies the characteristics of shock waves and this change is dependent also on the ionisation degree of the plasma. Dispersion can create short lived oscillations in the shocked plasma. The shock fronts become wider with the increase in the number of neutrals in the plasma.

\end{abstract}


\end{frontmatter}

\parindent=0.5 cm

\section{Introduction}

The solar atmosphere is an ideal medium for the propagation of disturbances (waves and oscillations), and these are channeled in most of the cases along the magnetic field lines that emerge from the solar interior and fill up the whole corona. In their propagation these waves are damped and dissipate their energy, they can be converted into other kind of waves, they can steepen into nonlinear waves, etc. The characteristics of these waves is very much dependent on the medium in which they propagate; as a matter of fact the interaction between waves and their environment is bi-directional. When waves dissipate their energy, this can be transformed into heat modifying the characteristic of the plasma environment (temperature, density, pressure, opacity, etc.), but the same characteristics will determine the speed, amplitude, frequency of waves. 

Given the large distances they can travel from the solar surface, waves can easily steepen into nonlinear waves due to, e.g. gravitational stratification-driven drop of plasma density with height, instabilities or the non-equilibrium state of the plasma (thermal, ionisation, etc.). On the other hand, there are mechanism that tend to take away energy from the wave (physical damping, due to dissipative effects, resonant absorption, phase mixing, etc.), and as a result the amplitude of waves will become smaller. Similarly, waves can be dispersive due to geometrical (waveguides) or physical (e.g. Hall currents, Finite Larmor radius effects) dispersion, and, as a result, the waves' front become wider, practically dispersing their energy. Both of these effects have the same result, i.e. the energy of waves averaged over a volume of plasma decreases. 

When the nonlinear growth of the amplitude is balanced by dispersion, magnetic field lines support the propagation finite-width solitons propagating with finite and constant speed preserving their shape. In contrast, when the nonlinear growth of waves is balanced by dissipation, shock waves can develop, which are thin structures (of the order of the mean free path of particles) propagating faster than the local characteristic speed. Shock waves are used to explain the heating of solar chromospheric plasma by dissipating their kinetic energy (e.g. \citet{Orta03, Ulmschneider05}), they can accelerate particles (e.g. \citep{bell78, cuntz2004, suzuki04}), but also shock waves driven by the leakage of global oscillations in the upper solar atmosphere can drive type-I spicules (e.g. \citet{depontieu2007, murawski10}, etc.) 

The scientific interest in partially ionised plasmas has increased recently not only in solar plasmas (e.g. photospheric structures, spicules,
prominence), but also in interstellar medium, molecular clouds, the formation of proto-stellar discs, planetary magnetospheres/ionospheres, exoplanets atmospheres, etc. (for a detailed review see, e.g. \citet{ballester18}, and references therein).

In these environments temperatures are not high enough to ensure a full ionisation of the plasma, meaning that at all times various species of particles can co-exist, such as ions, neutrals and electrons. Strictly speaking in such a mixture one can also find excited atoms/ions and photons. However, these are more connected to rapidly changing plasmas, in the case of a non-equilibrium and non-elastic collisions between particles. Although only charged particles are capable of being guided along magnetic field lines, neutrals are kept in the system by collisions especially with ions. Through these collision the constituent particles can exchange energy and momentum showing a collective motion.

The mathematical description of partially ionised plasmas differs from the standard magnetohydrodynamics (MHD) framework, as the governing equation vary depending on the scale over which we would like to study an effect, and the threshold is determined by the ion-neutral collisional frequency (or the associated collisional time). Accordingly, for characteristic times that are longer than this collisional time the dynamics can be treated within the framework of a single fluid plasma. In this regime any change in the state of the plasma can be smoothed through collisions between the massive particles within a period. Here the values of physical quantities appearing in the system of equation describing the dynamics and the state of the plasma are a summation over all species of the mixture \citep{zaqar11}. If the dynamics occurs over timescales that are below the ion-neutral collisional time (or in other word, this is the high frequency regime), a two fluid description is needed, where charged particles (positive ions and negative electrons) are treated as single fluid that interact with neutrals. In our study we will restrict ourself to the one fluid model.

Partially ionized plasmas introduce physical effects which are not considered in fully ionized plasmas, for instance, Cowling's resistivity, isotropic thermal conduction by neutrals, heating due to ion/neutral friction, heat transfer due to collisions, charge exchange, ionization energy, etc., which are crucial to fully understand the behaviour of astrophysical plasmas in different environments. 

So far advances in the dynamics of partially ionised plasmas in the solar physics context has been reached mainly in the linear theory. { \citet{forteza07} studied the propagation of linear waves in a single fluid partially ionised plasma in the presence of collisions between particles and obtained that fast magnetoacoustic waves are the most affected waves and the damping of these waves takes place in plasmas that are characterised by strong field, low density and ionisation factor.  The same problem but in the presence of plasmas flows, often observed in solar prominence, was discussed later by \citet{carbonell2010} where they determined that the presence of flow strongly influences the spatial damping of linear MHD waves, and the obtained damping of slow waves matches the attenuation characteristics of observed slow waves, while in the case of fast and Alfv\'en waves efficient damping was obtained for periods that are not currently observed in prominences. An interesting approach towards understanding the properties of waves in partially ionised plasmas is the study by \citet{ballester18a}, where the authors considered the changes in the wave properties during the process of heating and/or cooling in a optically thin plasma, i.e. the periods, damping times of waves and cut-off wavenumbers become time-dependent. Another important problem discussed within the framework of single-fluid MHD was the topic of instabilities. \citet{diaz2014} considered the Rayleigh-Taylor instability (RTI) in single fluid plasma in the case of surface waves propagating along a tangential discontinuity in the presence of gravity. They found that the interface is always unstable due to the presence of neutrals, however the collision between particles can diminish the growth rate of waves. Later, \citet{khomenko2014} analysed numerically the non-linear evolution of the RTI and compared the results in the fully ionised and partially ionised case. They found that in the partially ionised case the plasma is always unstable and the growth rate for partially ionised plasmas is 50\% larger than in the fully ionised case. The same problem of RTI was later analysed by \citet{ruderman18}, however they allowed the magnetic field to have a shear with respect to the direction of wave propagation. Their results show that the RTI becomes sensitive to the plasma ionisation degree only for plasmas corresponding to small values of plasma-beta and predominantly neutral plasma. Their determined that the RTI  develops only for wavenumbers that are below a cut-off value. The possibility of developing of the dissipative instability (connected to negative energy waves) was studied by \citet{ballai15} in incompressible limit and \citet{mather18} for compressible plasma. They considered the interface between a partially ionised prominence plasmas and a fully ionised and viscous corona. They obtained that for flow speeds that are below the Kelvin-Helmholtz instability (KHI) waves become unstable and the two dissipative mechanisms have different effect on the stability of the interface. They also applied the same model to study the evolution of the unstable interface between prominences and dark plumes, however the growth rate they obtained was too small to explain the observed evolution of these structures. }

{A completely new regime describing the dynamics of plasma can be obtained in two-fluid description, where now frequencies of interest are larger than the ion-neutral collisional frequencies. \citet{shaikh10} investigated the propagation of Alfv\'en waves in this environment and found that the propagation speed of these waves is diminished due to the collision between ions and neutrals, and therefore an efficient energy transfer takes place between the charged and neutral fluid. \citet{zaqar11} studied the properties of the whole wave spectrum in two-fluid plasmas. As expected, the properties of waves at low frequency do not differ significantly in from the ones obtained in a single fluid plasma. However, for higher frequencies, Alfv\'en and fast waves attain their maximum damping rate at values in excess of the ion-neutral collisional frequency. Due to the presence of neutrals the spectrum of waves becomes very rich. The problem of resonant energy transfer in inhomogeneous plasmas through the resonant absorption of Alfv\'en waves in cold plasma approximation was investigated by \citet{soler12a}. These authors compared the damping of waves due to resonant absorption and ion-neutral collisions and found that the resonant absorption, requiring small length scales, is much more efficient mechanism to damp kink waves in the solar chromosphere that collision between particles. The two-fluid approach was also used to study the onset and evolution of instabilities. \citet{soler12b} studied the linear evolution of the KHI in both, incompressible and compressible plasma at an interface. They found that in the incompressible limit waves are unstable for any flows, while in the incompressible limit the instability threshold depends on the density ratio of plasmas and ion-neutral collisional frequency. They also found that, contrary to expectations, for high density ratio the flow threshold for unstable behaviour is sub-Alfv\'enic and these flows can trigger instabilities due to the coupling between ions and neutrals. The problem of RTI in two-fluid approximation was studied by \citet{diaz2012} and they found that compressibility and the ion-neutral collisions can lower the growth rate of waves. Based solely on the growth rate of waves, they concluded that the RTI could explain the existence  pf prominence fine structure with life-time of the order of 30 minutes. \citet{martinez2017} used a five-fluid model (including electrons, hydrogen and helium ions and neutrals) to study the properties of waves, their damping and dispersion (due to Hall currents). Finally, the current understanding of the dynamics in partially ionised space plasmas (including Earth's atmosphere, solar atmosphere, molecular clouds, accretion disks) was reviewed recently by \citet{ballester18}.}

In contrast so far little progress was achieved in the field of nonlinear waves, and shocks, in particular. The development and propagation of shocks in these plasmas is of special interest since one of the key ingredients in the appearance of shocks are exactly the dissipation mechanisms that take into account the ionisation degree of the plasma. The combination of shocks and partial ionisation is also important because the dissipation of shocks can lead to an increased temperature behind the shock that can contribute to an additional ionisation of the plasma.

Our primary aim is to investigate the interplay between nonlinearity, dissipation and dispersion in a partially ionised plasma, in particular to analyse the effects of weak dispersion due to Hall currents on the characteristics of shocks waves propagating in a single fluid partially ionised plasma. Our paper is structured in the following way: in Section 2 we introduce the model and the mathematical formalism needed to describe the dynamics of waves and discuss the importance of dispersive and dissipative effects in the lower solar atmosphere. In Section 3 we derive the evolutionary equation of nonlinear waves propagating in this environment using a multiple scale method. Next, in Section 4 we will assume that dispersion is weak, and write solution in the form of series expansion, where the effect of dispersion will be given as a second order correction. The analytical solutions will be represented graphically and studied for varying amount of neutrals, inclination angle of the magnetic field and the strength of the magnetic field. Finally, our results are summarised in Section 5.

\section{Model and mathematical background}
  
We model the solar lower atmospheric plasma as a non-ideal partially ionised gas permeated by an uni-directional magnetic field situated in the $xz$ plane, so the equilibrium magnetic field has the form ${\bf B}_0=(B_0\sin\alpha, 0, B_0\cos\alpha)$, where $\alpha$ is the angle between the $z$ axis and the direction of the magnetic field. The partially ionised plasma is made up from positive ions, negative electrons and neutrals. The total density is $\rho=\rho_e+\rho_i+\rho_n\approx \rho_i+\rho_n$. The relative densities of neutrals and ions are defined as \citep{forteza07}
\begin{equation}
\xi_i=\frac{\rho_i}{\rho}\approx \frac{n_i}{n_i+n_n}, \quad \xi_n=\frac{\rho_n}{\rho}\approx \frac{n_n}{n_i+n_n}.
\label{eq:1.1}
\end{equation}
The degree of ionisation can be characterised by the ionisation fraction defined as
\begin{equation}
{\mu}=\frac{1}{1+\xi_i}.
\label{eq:1.2}
\end{equation}
According to this definition, a fully ionised gas corresponds to ${\mu}=0.5$, while a neutral gas is described by ${\mu}=1$. The last two relations allow us to express the relative densities in terms of the ionisation degree as
\[
\xi_i=\frac{1}{ \mu}-1, \quad \xi_n=2-\frac{1}{ \mu}.
\]
The equations of non-ideal fully ionised plasma can be obtained by taking $\xi_n=0$ and $\xi_i=1$.

In the present study we neglect the effects of particle ionisation and recombination, and also assume a strong thermal
coupling between the species, which leads electrons, ions and neutrals having the same temperature (i.e. $T_e = T_i = T_n = T$). Therefore, the three component gas can be considered as a single fluid. 
Our aim is not to investigate the structure of the shock waves in partially ionised plasmas and the effect of partial ionisation on the formation and propagation of shocks.  Since we are going to limit ourselves on linear dissipation, we will neglect effects connected to the inertia of different particles. In addition, we consider that thermodynamic quantities (pressure, temperature) are relatively smooth functions of the spatial coordinates, i.e. the relative densities of neutrals and ions are constants. Therefore, we will restrict our model to transport mechanisms that arise in the induction equation. For simplicity we assume that gravitational stratification is neglected and all dissipative coefficients are constants. As a result, the dynamics of plasma can be described by the system
\begin{equation}
\frac{D\rho}{Dt}+\rho\nabla\cdot {\bf v}=0,
\label{eq:1.3}
\end{equation}
\begin{equation}
\rho\frac{D{\bf v}}{Dt}=-\nabla p+{\bf J}\times {\bf B},
\label{eq:1.4}
\end{equation}
\begin{equation}
\frac{\partial B}{\partial t}-\nabla\times ({\bf v}\times {\bf B})=\nabla\times \left[\eta{\bf J}+\eta_H\frac{{\bf B}\times {\bf J}}{|{\bf B}|} +\eta_A\frac{[({\bf J}\times {\bf B})\times {\bf B}]}{|{\bf B}|^2}\right],
\label{eq:1.4.1}
\end{equation}
\begin{equation}
{\bf J}=\frac{\nabla\times {\bf B}}{\mu_0},
\label{eq:1.5}
\end{equation}
\begin{equation}
\frac{\partial p}{\partial t}+{\bf v}\cdot \nabla p+\gamma p\nabla \cdot {\bf v}=(\gamma-1)Q
\label{eq:1.6}
\end{equation}
where $D/Dt=\partial/\partial t+{\bf v}\cdot$ is the advective operator, $\mu_0$ is the magnetic permeability of free space, $\rho$ and $p$ denote the mass density and pressure of the plasma, ${\bf B}$ is the magnetic field and ${\bf v}$ is the velocity {and $Q=\eta J_{\parallel}^2+\eta_A J_{\perp}^2$ is the energy lost due to ohmic dissipation, with $J_{\parallel}$ and $J_{\perp}$ being the components of the current, ${\bf J}$, parallel and perpendicular to the ambient field.} The magnetic field has to satisfy the solenoidal condition, i.e. $\nabla \cdot {\bf B}=0$. The perturbed magnetic field and velocity are ${\bf b}=(b_x, b_y, b_z)$ and ${\bf v}=(v_x, v_y, v_z)$, respectively. 

The three terms on the right-hand side of the induction equation (\ref{eq:1.5}) are the classical ohmic term, the Hall term and the ambipolar or neutral term. The diffusion coefficients of these terms above (in the case of real diffusion, i.e. ohmic and ambipolar diffusion) depend on the collisional frequency, on the magnetic field strength and on the neutral fraction. While the classical resistivity diffuses field-aligned currents, the ambipolar diffusion is responsible for the diffusion of currents in the perpendicular direction to the ambient magnetic field. We should keep in mind that out of the three coefficients the Hall coefficient does not describe dissipation, instead this term gives rise to dispersion. 

Assuming a hydrogen plasma these coefficients are defined as
\[
\eta=\frac{m_e(\nu_{ei}+\nu_{en})}{e^2n_e\mu_0}, \quad \eta_H=\frac{|B|}{en_e\mu_0}=\frac{v_A^2}{\omega_i}, \quad \eta_A=\frac{\xi_n^2v_A^2}{\nu_{in}+\nu_{en}},
\]
where $\omega_i=eB_0/m_i$ is the ion cyclotron frequency. The collisional frequency between particles are given by (see, e.g. \citet{spizeri62, braginskii65})
\[
\nu_{in}=n_n\sigma_{in}\left(\frac{8k_BT}{\pi m_{in}}\right)^{1/2}, \quad \nu_{en}=n_n\sigma_{en}\left(\frac{8k_BT}{\pi m_{en}}\right)^{1/2}, 
\]
\[
\nu_{ei}=\frac{n_ee^4\Lambda}{3m_e^2\epsilon_0^2}\left(\frac{m_e}{2\pi k_BT}\right)^{3/2},
\]
where
\[
m_{in}=\frac{m_im_n}{m_i+m_n}\approx \frac{m_i}{2}, \quad m_{en}=\frac{m_em_n}{m_e+m_n}\approx m_e,
\]
and 
\[
\Lambda=23.4-1.15\log_{10}n_e+3.45 \log_{10}T,
\]
is the Coulomb logarithm and all number densities are given in cgs units and temperature, $T$, in $eV$. The collisional cross sections are constants and they take the value $\sigma_{in}=5\times 10^{-19}$ m$^2$ and $\sigma_{en}=10^{-19}$ m$^2$, respectively {(\citet{oster61,vranjes2013})}. Under solar atmospheric conditions and assuming a monoatomic (hydrogen) plasma the ratio between the i-n and e-n collisional frequencies is proportional to the small quantity $(m_e/m_i)^{1/2}$, therefore it is of the order of ${\cal O}(10^{-2})$. As a consequence, the e-n collisional frequency in the expressions of the ambipolar diffusion can be neglected. Furthermore, given that we are dealing with a hydrogen plasma $n_i=n_e$.


{The variation of the Hall coefficient, the coefficient of ambipolar resistivity and the coefficient of Spitzer resistivity with height was analysed by \citet{khom12} assuming VAL III C model \citep{vernazza81}. According to this study} the classical ohmic term reaches its largest values at photospheric heights, between 0 and 500 km. However, even there, the Hall term is about one order of magnitude larger than the ohmic term. The ambipolar term becomes dominant over the other two from 900 km upwards. At chromospheric heights, this term is up to 5 orders of magnitude larger than the ohmic term. 
In many studies it is customary to replace the ambipolar diffusivity by the Cowling diffusivity according to $\eta_A=\eta_C-\eta$, as for homogeneous magnetic fields this choice would cancel the ohmic dissipative term in the induction equation.  

We consider small but finite amplitude perturbations about a uniform equilibrium, writing all physical quantities $f$ in the form
\[
f=f_0+{\hat f},
\]
where $f_0$ is an equilibrium value and ${\hat f}$ its Eulerian perturbation. The {\it hat} symbol will be omitted subsequently. We suppose wave propagation to be along the z-axis, and we choose $\partial/\partial x =
\partial/\partial y = 0$. As a result, the system of MHD equations describing the evolution of nonlinear wavers in partially ionised plasma are given by
\begin{equation}
\frac{\partial \rho}{\partial t}+\rho_0\frac{\partial v_z}{\partial z}={\cal N}_1,
\label{eq:1.7}
\end{equation}
\begin{equation}
\rho_0\frac{\partial v_x}{\partial t}-\frac{B_0\cos\alpha}{\mu_0}\frac{\partial b_x}{\partial z}={\cal N}_2,
\label{eq:1.8}
\end{equation}
\begin{equation}
\rho_0\frac{\partial v_y}{\partial t}-\frac{B_0\cos\alpha}{\mu_0}\frac{\partial b_y}{\partial z}={\cal N}_3,
\label{eq:1.9}
\end{equation}
\begin{equation}
\rho_0\frac{\partial v_z}{\partial t}+\frac{\partial p}{\partial z}+\frac{B_0\sin\alpha}{\mu_0}\frac{\partial b_x}{\partial z}={\cal N}_4,
\label{eq:1.10}
\end{equation}
\begin{equation}
\frac{\partial b_x}{\partial t}+B_0\frac{\partial}{\partial z}(v_z\sin\alpha-v_x\cos\alpha)={\cal N}_5+{\cal D}_1+H_1,
\label{eq:1.11}
\end{equation}
\begin{equation}
\frac{\partial b_y}{\partial t}-B_0\cos\alpha\frac{\partial v_y}{\partial z}={\cal N}_6+{\cal D}_2+H_2,
\label{eq:1.12}
\end{equation}
\begin{equation}
\frac{\partial b_z}{\partial t}={\cal D}_3+H_3,
\label{eq:1.14}
\end{equation}
\begin{equation}
\rho_0\frac{\partial p}{\partial t}-\rho_0c_S^2\frac{\partial \rho}{\partial t}={\cal N}_7,
\label{eq:1.15}
\end{equation}
where the quantities ${\cal N}_i$ are nonlinear terms and ${\cal D}_i$ are non-ideal terms, respectively. Since $b_z$ does play any role in the above equation, we choose $b_z=0$. {We should note here that despite the sink terms in the energy equation (\ref{eq:1.6}), in the above equations we restricted ourself to linear dissipative terms. The energy lost due to the joule (ohmic) dissipation of parallel and perpendicular currents can appear in higher order (cubic) nonlinear approximations.} 

Restricting our analysis to quadratic nonlinearity, the nonlinear and dissipative terms in Eqs. (\ref{eq:1.7})--(\ref{eq:1.15}) are given by
\begin{equation}
{\cal N}_1=-\frac{\partial}{\partial z}(\rho v_z), \quad {\cal N}_2=-\rho\frac{\partial v_x}{\partial t}-\rho v_z\frac{\partial v_x}{\partial z},
\label{eq:1.16}
\end{equation}
\begin{equation}
{\cal N}_3=-\rho\frac{\partial v_y}{\partial t}-\rho_0 v_z\frac{\partial v_y}{\partial z}, 
\label{eq:1.17}
\end{equation}
\begin{equation}
{\cal N}_4=-\rho\frac{\partial v_z}{\partial t}-\rho_0 v_z\frac{\partial v_z}{\partial z}-\frac{b_x}{\mu_0}\frac{\partial b_x}{\partial z}-\frac{b_y}{\mu_0}\frac{\partial b_y}{\partial z},
\label{eq:1.18}
\end{equation}
\begin{equation}
{\cal N}_5=-\frac{\partial (v_zb_x)}{\partial z}, \quad {\cal N}_6=-\frac{\partial (v_zb_y)}{\partial z},
\label{eq:1.19}
\end{equation}
\begin{equation}
{\cal N}_7=-\rho\frac{\partial p}{\partial t}-\rho_0v_z\frac{\partial p}{\partial z}+\gamma p\frac{\partial \rho}{\partial t}+\gamma p_0v_z\frac{\partial \rho}{\partial z},
\label{eq:1.20}
\end{equation}
\begin{equation}
{\cal D}_1=\eta\frac{\partial^2b_x}{\partial z^2}+\frac{\eta_C-\eta}{|{\bf B}_0|^2}B_0^2\frac{\partial^2b_x}{\partial z^2},
\label{eq:1.21}
\end{equation}
\begin{equation}
{\cal D}_2=\eta\frac{\partial^2b_y}{\partial z^2}+\frac{\eta_C-\eta}{|{\bf B}_0|^2}B_0^2\cos^2\alpha\frac{\partial^2b_y}{\partial z^2}.
\label{eq:1.22}
\end{equation}
{The value of ${\cal D}_3$ remains unspecified as $b_z$ was chosen to be zero and this term is not required in the calculations.} In addition the Hall terms appearing in the components of the induction equation are defined as
\begin{equation}
H_1=-\frac{v_A^2}{\omega_i}\cos\alpha\frac{\partial^2 b_y}{\partial z^2}, \quad H_2=\frac{v_A^2}{\omega_i}\cos\alpha\frac{\partial^2 b_x}{\partial z^2},
\label{eq:1.23}
\end{equation}
The above equations will be used in the forthcoming calculations in order to determine the nonlinear equation describing the spatial and temporal evolutions of waves.

\section{Determination of the nonlinear evolutionary equation}

In order to obtain an evolutionary equation for nonlinear waves propagating in a partially ionised prominence plasma, we use the well-known method of multiple scaling. The two basic parameters which describe the solution are the measure of the strength of nonlinearity given by the ratio $\Delta \rho/\rho_0$ and the measure of the strength of dissipation, $D_d$, due to partial ionisation. A shock wave is formed when these two effects are balanced in magnitude. Let us consider these effects small and of the same order, i.e.
\begin{equation}
\Delta \rho/\rho_0\sim D_d\sim\epsilon\ll 1,
\label{eq:2.1}
\end{equation}
where the quantity $\Delta \rho$ is the growth in the amplitude due to nonlinearity. Their forms remain unspecified as they are not used in our analysis. 

Solving the MHD equations and looking for wave-like solutions we would obtain that the solutions involve two characteristic times. One is the time that is associated with the oscillatory motion (i.e. period of waves) and the other time that is involved in the damping of waves due to dissipation. Since we are interested in the weak damping of plane propagating waves, these two times will be comparable if we introduce the "slow time", $\tau$. Therefore, according to the standard method {of multiple scaling (see, e.g. \citet{kevor})}, we introduce the stretched variables $\xi$ and $\tau$ through
\begin{equation}
\xi=\epsilon^{1/2}(z-Vt), \quad \tau=\epsilon^{3/2}t,
\label{eq:2.2}
\end{equation}
where $V$ is the phase velocity of a wave (slow or fast magnetoacoustic). Next, we expand all perturbations in power series, so that the $x$ and $z$ components of the velocity and magnetic field perturbations, density, pressure take the form
\begin{equation}
g(z,t)=\epsilon g_1(z,t)+\epsilon^2g_2(z,t)+\dots,
\label{eq:2.3}
\end{equation}
while the Alfv\'enic variables (the $y$ components of the velocity and magnetic field perturbation) will be expanded according to
\begin{equation}
{h}(z,t)=\epsilon^{1/2}[\epsilon {h}_1(z,t)+\epsilon^2{h}_2(z,t)+\dots].
\label{eq:2.4}
\end{equation}
With these particular type of expansions we ensure that the Alfv\'enic fluctuations are $\epsilon^{1/2}$ smaller than the compressional variables and allow us to isolate compressional waves despite the coupled equations in the MHD system of equations. Our aim is to derive an equation where nonlinearity, cubic dispersion and quadratic dissipation are of the same order. {In this limit the dispersion will not affect the linear waves, together with dissipation, dispersion is a second order effect.}

Let us consider that the characteristic length along the $z$-axis is $l_{ch}$. From the theory of partial differential equation we know that terms with odd order derivatives describe dispersion, while terms with even order order derivatives will describe dissipation. Our aim here is to derive an equation when all three effects (nonlinearity, dispersion, and dissipation) are of the same order. The ratio between a typical nonlinear and typical dispersive term is of the form 
\[
\frac{g\;\partial g/\partial z}{\partial^3 g/\partial z^3},
\]
where $g(z,t)$ is any variable that has an expansion given by Eq.  (\ref{eq:2.3}). Since this ratio has to be of order one, it is easy to show that in the leading order $\epsilon l_{ch}^2={\cal O}(1)$. Next the typical ratio of the nonlinear term and the dissipative term is
\[
\frac{g\;\partial g/\partial z}{\eta_C\partial^2 g/\partial z^2}.
\]
It is easy to show that this ratio will be of the order of 1 if $\eta_C\epsilon^{-1/2}={\cal O}(1)$, which suggests that the dissipative coefficients ($\eta$ and $\eta_C$) should be rescaled as
\begin{equation}
\eta=\epsilon^{1/2}{\bar \eta}.
\label{eq:2.5}
\end{equation}
After introducing the new set of variables into the MHD equations and collecting terms of subsequent powers of $\epsilon$, we obtain a sequence of equations for variables $g(z,t)$ and ${h}(z,t)$ corresponding to different powers of the small parameter $\epsilon$. In the lowest order of approximation (terms proportional to $\epsilon^{3/2}$) we recover a set of equations describing the propagation of linear magnetoacoustic waves in an ideal plasma, and these equations are similar to the equations we would obtain for a fully ionised case. This means that partial ionisation is a secondary effect and together with nonlinearity will affect waves only in the higher approximations. In the lowest approximation the equations can be easily obtained by setting all nonlinear, dissipative and dispersive terms in (\ref{eq:1.7})--(\ref{eq:1.15}) to zero. Consequently all variables can be expressed in terms of the perturbed density ($\rho_1$) as
\begin{equation}
v_{z1}=\frac{V}{\rho_0}\rho_1,\quad p_1=c_S^2\rho_1,
\label{eq:2.6}
\end{equation}
\begin{equation}
v_{x1}=-\frac{(V^2-c_S^2)\cot \alpha}{\rho_0V}\rho_1, 
\label{eq:2.7}
\end{equation}
\begin{equation}
 b_{x1}=B_0\frac{V^2-c_S^2}{\rho_0v_A^2\sin\alpha}\rho_1,
\label{eq:2.8}
\end{equation}
where the phase speed, $V$, satisfies the linear dispersion relation
\begin{equation}
V^4-V^2(c_S^2+v_A^2)+c_S^2v_A^2\cos^2\alpha=0,
\label{eq:2.9}
\end{equation}
and $v_A=B_0/(\mu_0\rho_0)^{(1/2)}$ and $c_S=(\gamma p_0/\rho_0)^{(1/2)}$ are the Alfv\'en and sound speeds, respectively. Relation (\ref{eq:2.9}) means that linear fast and slow magnetoacoustic waves will propagate with a phase speed
\begin{equation}
V_{f,s}^2=\frac12\left(c_S^2+v_A^2\pm\sqrt{c_S^4+v_A^4-2c_S^2v_A^2\cos 2\alpha}\right),
\label{eq:2.9.1}
\end{equation}
where the indices $f$ and $s$ stand for fast and slow wave, respectively. Finally, the Alfv\'enic perturbations are connected through
\[
b_{y1}=-\frac{B_0V}{va_A^2\cos\alpha}v_{y1},
\]
and the $y$-component of the induction equation results in 
\begin{equation}
v_{y1}=\frac{1}{\omega_i}\left(\frac{V^2-c_S^2}{V^2-v_A^2\cos^2\alpha}\right)v_A^2\cos^2\alpha\frac{\partial}{\partial \xi}\left(\frac{ \rho_1}{\rho_0}\right).
\label{eq:2.9.2}
\end{equation}

In the next order of approximation corresponding to terms proportional to $\epsilon^{5/2}$ we obtain a system of equation for variables with an index $2$ that are expressed in terms of variables in the previous order of approximation. In this approximation the equations that will be used can be written as
\begin{equation}
V\frac{\partial \rho_2}{\partial \xi}-\rho_0\frac{\partial v_{z2}}{\partial \xi}=\frac{V}{\rho_0}\frac{\partial \rho_1^2}{\partial \xi}+\frac{\partial \rho_1}{\partial \tau},
\label{eq:2.9.3}
\end{equation}
\begin{equation}
\rho_0V\frac{\partial v_{x2}}{\partial \xi}+\frac{B_0\cos\alpha}{\mu_0}\frac{\partial b_x}{\partial \xi}=-\frac{(V^2-c_S^2)\cot \alpha}{V}\frac{\partial \rho_1}{\partial \tau},
\label{eq:2.9.4}
\end{equation}
\begin{equation}
\rho_0V\frac{\partial v_{z2}}{\partial \xi}-\frac{\partial p_2}{\partial \xi}-\frac{B_0\sin\alpha}{\mu_0}\frac{\partial b_{x2}}{\partial \xi}=V\frac{\partial \rho_1}{\partial \tau}+\frac{(V^2-c_S^2)^2}{\rho_0v_A^2\sin^2\alpha^2}\rho_1\frac{\partial \rho_1}{\partial \xi},
\label{eq:2.9.5}
\end{equation}
\[
V\frac{\partial b_{x2}}{\partial \xi}-B_0\frac{\partial}{\partial \xi}\left(\sin\alpha v_{z2}-\cos\alpha v_{x2}\right)=\frac{B_0(V^2-c_S^2)}{\rho_0v_A^2\sin\alpha}\frac{\partial \rho_1}{\partial \tau}+\frac{VB_0(V^2-c_S^2)}{\rho_0^2v_A^2\sin\alpha}\frac{\partial \rho_1^2}{\partial \xi}+
\]
\begin{equation}
+{\bar \eta}_C\frac{B_0(V^2-c_S^2)}{\rho_0v_A^2\sin\alpha}\frac{\partial^2\rho_1}{\partial \xi^2}-\frac{B_0Vv_A^2\cos^2\alpha}{\rho_0\omega_i^2}\left(\frac{V^2-c_S^2}{V^2-v_A^2\cos^2\alpha}\right)\frac{\partial^3\rho_1}{\partial \xi^3},
\label{eq:2.9.6}
\end{equation}
\begin{equation}
\frac{\partial p_2}{\partial \xi}-c_S^2\frac{\partial \rho_2}{\partial \xi}=\frac{c_S^2}{\rho_0}(\gamma-1)\rho_1\frac{\partial \rho_1}{\partial \xi}.
\label{eq:2.9.7}
\end{equation}

The compatibility condition imposed on this system results in the evolutionary equation
\begin{equation}
\frac{\partial \rho_1}{\partial \tau}+\alpha_1\rho_1\frac{\partial \rho_1}{\partial \xi}-\alpha_2\frac{\partial^2 \rho_1}{\partial \xi^2}-\alpha_3\frac{\partial^3 \rho_1}{\partial \xi^3}=0,
\label{eq:2.10}
\end{equation}
which is a Korteweg-de Vries-Burgers (KdVB) equation describing the propagation of a nonlinear wave in a dissipative and dispersive plasma. In the above equation the coefficients $\alpha_1$, $\alpha_2$ and $\alpha_3$ are given by e.g. \citet{ballai03}
\[
\alpha_1=V\frac{3V^2(V^2-c_S^2)+c_S^2(\gamma+1)(V^2-v_A^2\cos^2\alpha)}{2\rho_0(V^4-c_S^2v_A^2\cos^2\alpha)},
\]
\[
\alpha_2=V^2\frac{{\bar \eta}_C(V^2-c_S^2)}{2(V^4-c_S^2v_A^2\cos^2\alpha)},
\]
\begin{equation}
\alpha_3=\frac{v_A^4\cos^2 \alpha
V^3(V^2-c_S^2)}{2\omega_i^2(V^2-v_A^2
\cos^2\alpha)(2V^4-v_A^2c_S^2\cos^ 2\alpha)}. \label{eq:2.27}
\end{equation}
Equation (\ref{eq:2.10}) describes the nonlinear evolution of both slow and fast magnetoacoustic waves in a dissipative plasma, but not their coupling (provided $V$ is replaced by the corresponding value). The second coefficient of Eq. (\ref{eq:2.10}) depends not only on the particular values of the characteristic speeds and the propagation angle but also on the dissipative coefficient, i.e. on the ionisation degree of the plasma.  It is interesting to note that despite we considered the effect of both, classical and Cowling resistivities, the damping of shock waves is due only to perpendicular currents, i.e. only the Cowling resistivity will affect the shock wave. According to the standard definition, for $\alpha=0$ we have parallel (slow and fast) shocks, while in the $\alpha=\pi/2$ case only perpendicular fast shocks will be able to propagate. In all other cases the shock wave is oblique. In the above equation $\omega_i=eB_0/m_i$ is the ion cyclotron frequency. In the case of fast magnetoacoustic waves all three coefficients are positively defined for all values of inclination angle, characteristic speeds and ionisation degree.

\section{Dispersive shock waves}

The nonlinear evolutionary equation derived earlier (Eq. \ref{eq:2.10}) is a nonlinear equation that describes the spatial and temporal evolution of different kind of waves, depending whether dissipation and/or dispersion is dominant. When the nonlinear growth of the amplitude is balanced by dispersion (here the last term in Eq. \ref{eq:2.10}) the evolutionary equation describes the KdV soliton (see, e.g. \citet{ballai03}). In this limit the evolutionary equation does not contain information about the partially ionised character of the plasma, therefore this limit will not be discussed. On the other hand if the nonlinear steepening of waves is balanced by a damping (here described by the second order derivative term), the governing equation describes shock waves propagating in partially ionised plasma. Here information on the ionisation degree of the plasma is included in the dissipative coefficient. 
In the present paper we will consider all three effects (nonlinearity, dispersion, dissipation) but the effect of dispersion due to Hall current will be considered week compared to dissipation.  
Based on the estimations by \citet{khom12} this approximation is valid for regions corresponding to heights $z>900$ km in the solar chromosphere. 

To obtain the governing equations for dispersive shocks propagating in partially ionised plasmas we follow closely the method developed by \citet{ballai07}. Accordingly, the effect of dispersion is going to be considered weak and, therefore, a correction to the solution of the Burgers equation (which is obtained by choosing $\alpha_3=0$ in Eq. \ref{eq:2.10}). In Burgers equation the effect of nonlinear steepening of waves is balanced by dissipation. The stationary solution of the Burgers equation in terms of the hyperbolic function is well know and is given by
\begin{equation}
\rho=\frac{N}{\alpha_1}\left[1-\tanh(\theta)\right], \label{eq:2.28}
\end{equation}
where the argument $\theta$ is written as
\[
\theta=\frac{N}{2\alpha_2}(\xi-N\tau),
\]
and $N$ is the speed of the shock wave in the system moving
together with the shock.

Since we assume weak dispersion, in Eq. (\ref{eq:2.10}) we consider that
$1\sim\alpha_1\sim\alpha_2\gg \alpha_3$. Let us apply a
reductive perturbative method for Eq. (\ref{eq:2.10}). The density
is expanded in terms of the small parameter, $\epsilon$ measuring
the dimensionless amplitude of waves and write
\begin{equation}
\rho=\epsilon\rho_1+\epsilon^2\rho_2+\epsilon^3\rho_3+\dots
\label{eq:3.1}
\end{equation}
In the lowest order in $\epsilon$ we obtain the Burgers equation describing the evolution of shocks with solutions given by Eq. (\ref{eq:2.28}) that satisfies the boundary conditions $\xi\to\pm \infty$
\begin{equation}
{\cal L}_0\rho_1=\frac{\partial \rho_1}{\partial
\tau}+\alpha_1\rho_1\frac{\partial}\rho_1{\partial
\xi}-\alpha_2\frac{\partial^2\rho_1}{\partial \xi^2}=0, \label{eq:3.2}
\end{equation}
where  ${\cal L}_0$ is the so called Burgers operator.

In the second order approximation ($\sim O(\epsilon^4)$), Eq.
(\ref{eq:2.10}) can be written as
\begin{equation}
{\cal L}_1\rho_2=\alpha_3\frac{\partial^3\rho_1}{\partial \xi^3},
\label{eq:3.5}
\end{equation}
where the linearized Burgers operator ${\cal L}_1$ is defined as
\begin{equation}
{\cal L}_1=\frac{\partial }{\partial \tau}+\alpha_1\frac{\partial
(\rho_1\cdot)}{\partial \xi}-\alpha_2\frac{\partial^2}{\partial
\xi^2}. \label{eq:3.6}
\end{equation}
The particular type of expansion used in the present study implies that dispersion is a secondary effect and it appears only in higher order approximation as a correction to shock waves. 
Substituting the solution obtained in the first order
approximation into Eq. (\ref{eq:3.5}) we obtain
\begin{equation}
{\cal L}_1\rho_2=\frac{\alpha_3N^4}{4\alpha_1\alpha_2^3}\left[3\sech^4(\theta)-2\sech^2(\theta)\right].
\label{eq:3.7}
\end{equation}

The two terms on the RHS of Eq. (\ref{eq:3.7}) can lead to secular growth of the solution. Indeed, it can
be easily shown that a solution proportional to $\sech^2(\theta)$ of the
equation ${\cal L}_1\rho_2=0$ leads to a secular behaviour in time
of the form $\tau \sech^2(\theta)$. In the case of the term proportional
to $\sech^4(\theta)$ it is necessary to examine the solution of the
equation ${\cal L}_1\rho_2={\cal C} \sech^4(\theta)$ where ${\cal C}$ is an
arbitrary constant. The solution of this equation will be
proportional to
\[
\sech^2(\theta)\ln \cosh(\theta)+\tanh^2(\theta).
\]
Due to the term $\tanh^2(\theta)$ this solution does not satisfy the
boundary condition at $\pm \infty$, therefore, we drop the term
$\tanh^2(\theta)$ and use an incomplete solution proportional to
$\sech^2(\theta)\ln \cosh(\theta)$ which shows, again, a secular behaviour.
The secularity in the solution of the equation in the second order
approximation can be easily removed using a renormalization technique introduced by \citet{kodama1978}.

Let us introduce the new variables
\[
{\tilde \xi}=\xi-\epsilon\delta_1\tau,
\]
\[
{\tilde \tau}=\tau,
\]
where the quantity $\delta_1$ is a renormalization constant. In
the new variables, the two operators are written as
\begin{equation}
{\tilde {\cal L}}_0=\frac{\partial }{\partial {\tilde
\tau}}+\alpha_1\rho_1\frac{\partial}{\partial {\tilde
\xi}}-\alpha_2\frac{\partial^2}{\partial {\tilde \xi}^2},
\label{eq:3.8}
\end{equation}
and
\begin{equation}
{\tilde {\cal L}}_1=\frac{\partial }{\partial {\tilde
\tau}}+\alpha_1\frac{\partial (\rho_1\cdot)}{\partial {\tilde
\xi}}-\alpha_2\frac{\partial^2}{\partial {\tilde \xi}^2}.
\label{eq:3.9}
\end{equation}
With these new notations, the equation obtained in the first order
approximation (see, Eq. \ref{eq:3.2}) is transformed into
\begin{equation}
{\cal L}_0\rho_1={\tilde {\cal
L}}_0\rho_1-2\epsilon\alpha_2\delta_1\frac{\partial^2
\rho_1}{\partial {\tilde
\xi}^2}-\epsilon^2\delta_1^2\alpha_2\frac{\partial^2
\rho_1}{\partial {\tilde \xi}^2}. \label{eq:3.10}
\end{equation}
By introducing the new variables, higher order terms in $\epsilon$
are generated. The term in Eq. (\ref{eq:3.10}) proportional to
$\epsilon$ will be added to the next order approximation, while
the term proportional to $\epsilon^2$ will be part of a higher
order approximation, not written explicitly here. Therefore, in the
first order approximation we obtain
\begin{equation}
{\tilde {\cal L}}_0\rho_1=0, \label{eq:3.11}
\end{equation}
which admits a similar solution as presented in (\ref{eq:2.28})
with $\theta$ replaced by ${\tilde \theta}$ and $\xi$ and $\tau$
replaced by ${\tilde \xi}$ and ${\tilde \tau}$.

In the next order of approximation the terms obtained in the
lowest order of approximation must be added, so now the equation
which must be solved is
\begin{equation}
{\tilde {\cal L}}_1\rho_2-2\alpha_1\delta_1\frac{\partial^2
\rho_1}{\partial {\tilde
\xi}^2}=\frac{\alpha_3\rho_0^4}{4\alpha_1\alpha_2^3}\left[3\sech^4(\theta)-2\sech^2(\theta)\right].
\label{eq:3.12}
\end{equation}
The secular terms on the right hand side of Eq. (\ref{eq:3.12})
are eliminated by a proper choice of the renormalization constant
$\delta_1$.

Assuming a solution of Eq. (\ref{eq:3.12}) in the form
\[
\rho_2=\delta_2\sech^2({\tilde \theta})\ln \cosh({\tilde \theta}),
\]
we can easily obtain that the renormalization constant and the
coefficient $\delta_2$ are given by
\begin{equation}
\delta_1=-\frac{\alpha_3\rho_0^2}{\alpha_2}, \quad
\delta_2=-\frac{3\alpha_3\rho_0^2}{\alpha_1\alpha_2^2},
\label{eq:3.14}
\end{equation}
With this renormalization constant the secular growth of solutions
is removed and the solution of the Burgers equation in the
presence of weak dispersion is given by
\begin{equation}
\rho=\epsilon\rho_1+\epsilon^2\rho_2+O(\epsilon^3), \label{eq:3.15}
\end{equation}
where the $\rho_1$ is given by Eq. (\ref{eq:2.28}) and
\[
\rho_2=-\frac{3\alpha_3N^2}{\alpha_1\alpha_2^2}\sech^2({\tilde
\theta})\ln\cosh({\tilde \theta}),
\]
and the argument of the function, ${\tilde \theta}$ is given by
\begin{equation}
{\tilde \theta}=\frac{N}{2\alpha_2}\left[\xi-\tau
N\left(1-\frac{\epsilon\alpha_3N}{\alpha_2^2}\right)\right].
\label{eq:3.16}
\end{equation}
This solution shows that the velocity of the shock wave is
modified by the presence of dispersion as
\begin{equation}
W=N\left(1-\frac{\epsilon\alpha_3N}{\alpha_2^2}\right).
\label{eq:3.17}
\end{equation}

It is instructive to analyse the percentage change of the velocity of the shock due to the consideration of weak dispersion for various angles of propagation and ionisation fraction. From Eq. (\ref{eq:3.17}) it is easy to find that 
\begin{equation}
\Delta=\frac{N-W}{N}=\epsilon\frac{\alpha_3N}{\alpha_2^2}.
\label{eq:3.18}
\end{equation}

Let us investigate the effect of partial ionisation on the evolution of these shocks. We consider the spatial and temporal evolution of shock waves that are obtained as a result of the steepening of fast magnetoacoustic waves (Figure 1).  We plot the spatial evolution of these shocks for three distinct times for an inclination angle of 20 degrees and an ionisation degree that is very close to the fully ionised limit.
\begin{figure}
\label{figure2}
\begin{center}
\includegraphics*[width=10cm]{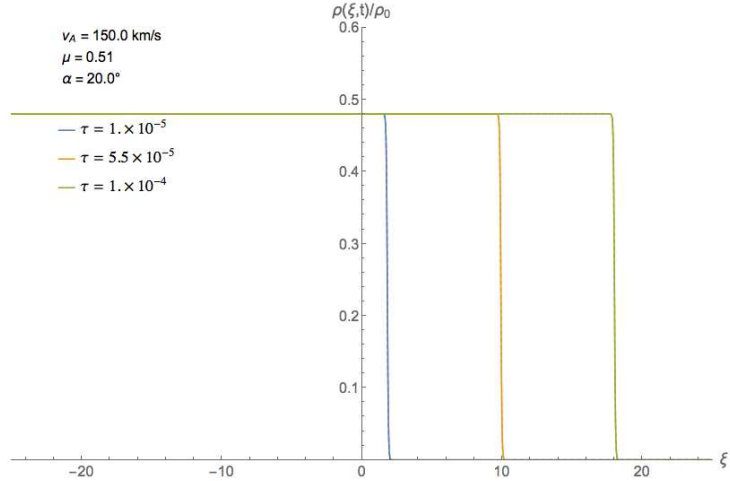}
\end{center}
\caption{The spatial evolution of a fast shock wave under the effect of weak dissipation. Here the three wavefronts correspond to three different times. In this plot the inclination angle is 20$^\circ$, the Alfv\'en speed is 150 km s$^{-1}$ (meaning that plasma beta is  $1.2\times 10^{-2}$) and the plasma is nearly fully ionised ($\mu=0.51$).}
\end{figure}

In this plot the solid lines represent the evolution of a pure shock wave whose solution is given by Eq. (\ref{eq:2.28}), and in all subsequent plots the solution that corresponds to the case of dispersive shock will be plotted by dashed lines. In the case of a plasma that is nearly fully ionised the effects of dissipation and dispersion are very weak; practically the solutions that correspond to these two cases overlap. The situation is different when the number of neutrals is increased (see Figure 2).
\begin{figure}
\label{figure3}
\begin{center}
\includegraphics*[width=10cm]{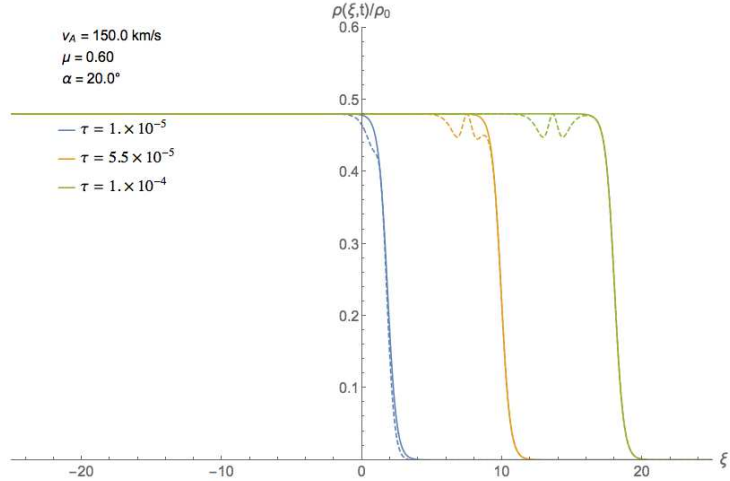}
\end{center}
\caption{The same as in Fig 1, but here the ionisation degree is $\mu=0.6$.}
\end{figure}

First of all we need to note that the presence of neutrals widens the shock front, and considering Figures 2-4 it is clear that the width of the shock waves increases with the amount of neutrals in the system. This effect is due to the fact that the Cowling resistivity increases due to the enhanced proportion of neutrals (see the earlier definition of coefficient of ambipolar, and consequently, the Cowling diffusion coefficient). The physical background of this effect resides in the way the ambipolar diffusion is defined.  With the increase of neutrals the collisional frequency between ions and neutrals increases  and this enhances the diffusion in the perpendicular direction to the magnetic field. The width of the shock front is proportional to the magnitude of the ambipolar (or Cowling) resistivity.
\begin{figure}
\label{figure4}
\begin{center}
\includegraphics*[width=10cm]{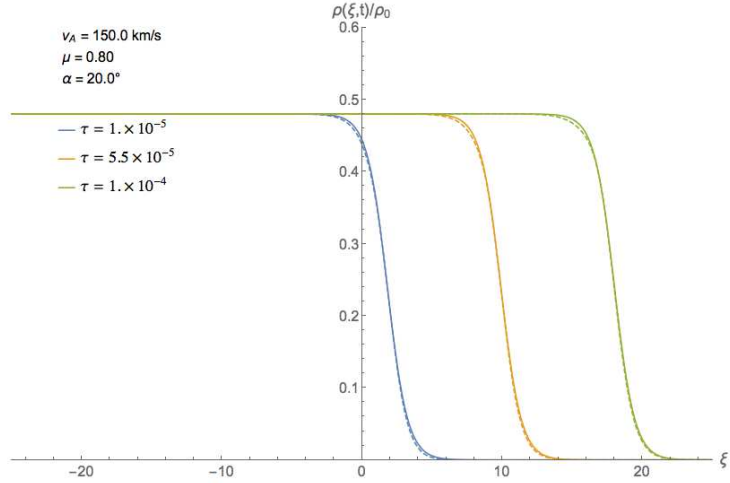}
\end{center}
\caption{The same as in Figure 1 but here the ionisation degree is $\mu=0.8$.}
\end{figure}

The dissipation (and, therefore, the weak dispersion) is relevant on a narrow region that constitute the thickness of the shock. Far away from this transition region the solutions obtained for dispersive and non-dispersive solutions should coincide. Another important consequence of the dispersion is the modified profile of the shock right after the shock. As a matter of fact this peculiar behaviour can be better seen and interpreted when we decrease the Alfv\'en speed (see Fig. 4). Now it is clear that the change in the shock front is in fact an short lived oscillatory behaviour of the post-shock density and it is relevant for lower values of the magnetic field. When the magnetic field is decreased, the radius of gyration of ions around the magnetic field is increased, therefore the gyro frequency is decreasing. As a result the inertial length of ions is increased. It is likely that because of decreased ion cyclotron frequency the dispersive effects are more important (the value of the dispersive coefficient increases) and that modifies the distribution of energy in the shock front to the level where short lived oscillatory motion can appear behind the shock. For a given magnetic field the oscillatory behaviour in the post-shocked plasma appears for a low concentration of neutrals (in Fig 2 two thirds of the plasma is fully ionised), however this is smoothed once the concentration of neutrals increases. 
\begin{figure}
\label{figure5}
\begin{center}
\includegraphics*[width=10cm]{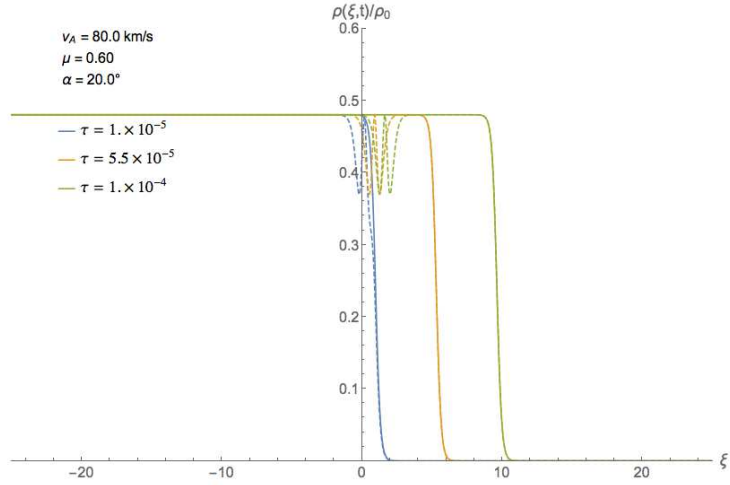}
\end{center}
\caption{The same as in Figure 1 but here the ionisation degree is $\mu=0.6$ and Alfv\'en speed is reduced to 80 km s$^{-1}$.}
\end{figure}

\begin{figure}
\label{figure6}
\begin{center}
\includegraphics*[width=10cm]{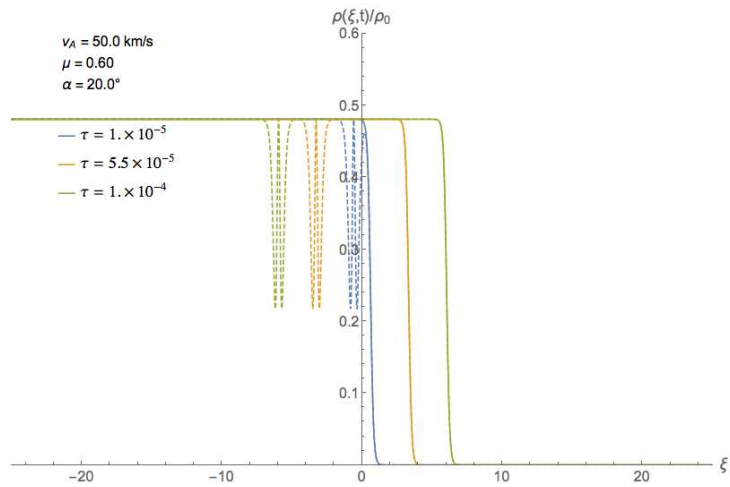}
\end{center}
\caption{The same as in Figure 1 but here the ionisation degree is $\mu=0.6$ and Alfv\'en speed is reduced to 50 km s$^{-1}$.}
\end{figure}

If we decrease even further the magnetic field (i.e. the Alfv'en speed), the amplitude of oscillations in the post-shocked plasma is even higher, as shown in Fig. 5. Since we are dealing with nonlinear waves, a decrease in the amplitude of waves would translate in the decrease of their propagation speed.
These figures where repeated for various inclination angles of the magnetic field and our numerical analysis reveal that given the almost isotropic phase speed of fast magnetoacoustic waves with the inclination angle, this is preserved even in the properties of nonlinear waves, i.e. the height of the shock and its speed does not vary significantly with the inclination of the magnetic field.  

\section{Conclusions}

The problem of shock wave formation, propagation and damping and the effects these shocks induce in their medium of  is one of the most important aspects of dynamics in solar plasmas, especially if we want to understand the heating in the lower solar atmosphere. In this region the plasma is not hot enough for a full plasma ionisation, therefore the dynamics of shock waves will be influenced by the presence of neutrals.

Our study dealt with the propagation characteristics of shocks waves in the presence of Cowling resistivity and weak dispersion caused by Hall currents. Due to the weakness of dispersion, this effect adds a second order correction to the shock solution of evolutionary equation. In this study we focussed on wave steepening in a plasma where the frequency of linear oscillations is much higher than the ion-neutral frequency, i.e. we used the single fluid description. 

The dynamics of shock waves is described by the Burgers equation, whose travelling wave solution is a well-known {\it tanh} function. When considering the effect of dispersion a re-normalization technique was used to avoid secular behaviour of dispersive terms in the solution. 

Our results show that the width of shocks increases with the amount of neutrals in the system. Dispersion reduces the height (and consequently the speed) of shock waves . This property must be connected to the character of dispersive linear fast magnetoacoustic waves showing a decreased propagation speed with the increase of dispersion, i.e. these waves have negative dispersion. On the contrary we expect that the shock waves that develop as a result of the nonlinear steepening of slow magnetoacoustic waves will show an opposite effect, the phase speed of these waves increases with the effect of dispersion, therefore they have a positive dispersion. 

It remains to be investigated what is the effect of the weak dispersion on the effectiveness of plasma heating by shock dissipation, and how these dispersive shocks can heat the chromospheric plasma.

\section{Acknowledgements}

I.B. was partially supported by the Leverhulme Trust (IN-2014-016). IB and AM were supported by a grant of the Romanian Ministry of National Education and Scientific Research, RDI Programme for Space Technology and Advanced Research - STAR (181/20.07.2017). Part of this study was carried out while I.B. was a Domus Hungarica Scientiarum et Artium fellow. Their support, and the hospitality of the staff at the Astronomy Department at E\"otv\"os Lor\'and University is greatly acknowledged.

\section{Citations}




\end{document}